\documentclass{webofc}
\usepackage[varg]{txfonts}   % Web of Conferences font

\usepackage{siunitx}
\begin{document}
\title{The strangeness program at GlueX}

\author{\firstname{Peter} \lastname{Pauli}\inst{1}\fnsep\thanks{\email{Peter.Pauli@glasgow.ac.uk}} , for the GlueX collaboration
}

\institute{School of Physics and Astronomy, G12 8QQ, Glasgow, UK}

\abstract{%
  The GlueX experiment located at Jefferson Lab studies the spectrum of hadrons using photoproduction on a LH$_2$ target in a wide variety of final states. With its detector system capable of measuring neutral and charged final state particles over almost the full solid angle, and very good particle identification capabilities, GlueX can measure many different hadrons containing strangeness. A linearly polarized photon beam allows the measurement of polarization observables, which contain information about the production mechanisms involved in generating strange particles in photoproduction. In addition, GlueX can perform precise cross-section measurements, which help to study the spectrum of strange hadrons. In this presentation, the GlueX experiment is introduced, and recent progress of its strangeness program is discussed. We present recent results on $\Sigma^0$ beam asymmetries, $\Lambda(1520)$ spin-density matrix elements and ongoing studies of the $\Lambda(1405)$ lineshape. We also present our recent progress on measurements of $\Lambda\bar\Lambda$ and $\Xi^{(*)}$ photoproduction. Also, future prospects for strangeness measurements at GlueX are discussed.
}
\maketitle
\section{Introduction}\label{sec:intro}
Although much progress has been made over the last years in studying the hadron spectrum of light and heavy quark states, the spectrum of hadrons containing strange quarks is only poorly understood. For example, at present only a fraction of the hyperon states predicted by theory are experimentally confirmed. One possibility to expand the available data is the study of strange hadrons in photoproduction. It provides a  complementary production process to hadron beam experiments and utilizing linearly polarized photons enables the study of production mechanisms.\par
GlueX is a photoproduction experiment located in Jefferson Lab's Hall D. A linearly polarized photon beam is produced off a diamond radiator via the coherent bremsstrahlung technique. The diamond is oriented such that the largest fraction of polarization is located at photon energies between \SIrange{8.2}{8.8}{\GeV}. This fixed target experiment uses a liquid hydrogen target which is surrounded by the GlueX detector setup, consisting of multiple tracking detectors and calorimeters as well as particle identification detectors providing almost 4$\pi$ acceptance. The main motivation behind the GlueX experiment is the search for exotic hybrid mesons. Since these are expected to decay into charged and neutral final state particles, GlueX was designed to have good energy and momentum resolution for both. A comprehensive description of the GlueX detector system and beamline can be found in Ref.~\cite{GlueX:2020idb}. The large acceptance paired with good resolutions for different particle types makes GlueX an ideal detector system to study strange hadrons, which often decay into final states with charged and neutral particles. We will show that GlueX has an exciting strangeness program with high statistical precision for many strange hadron reactions. \par
GlueX's first campaign of data taking (GlueX-I) ran from 2017-18 and resulted in an integrated luminosity of about \SI{440}{\pico\barn^{-1}} for $\SI{6}{\GeV}<E_\gamma<\SI{11.6}{\GeV}$. Afterwards GlueX was upgraded with a new DIRC detector to improve $\pi/K$ separation and GlueX-II is taking data since 2020. Unless otherwise stated the data presented in this talk represents the full GlueX-I statistics.

\section{$\Lambda\bar\Lambda$ production}\label{sec:LLbar}
GlueX is well suited to study baryon-anti-baryon photoproduction and some preliminary results for $p\bar p$ and $\Lambda\bar\Lambda$ have already been reported at previous conferences~\cite{Schumacher:2018xnh,Li:2019rts}. These reactions are interesting as they allow to search for baryonium, a bound baryon-anti-baryon system which has so far not been observed. However, BES~\cite{BES:2003aic} and BESIII~\cite{BESIII:2017hyw} reported interesting cross-section enhancements near threshold for $e^+e^-\rightarrow\gamma p\bar p$ and $e^+e^-\rightarrow\Lambda\bar\Lambda$, respectively. This motivated a closer examination of these two final states at GlueX, where these reactions can be measured with high statistical accuracy over a wide range of photon beam energies.\par
\begin{figure}[ht]
    \centering
    \includegraphics[height=4cm]{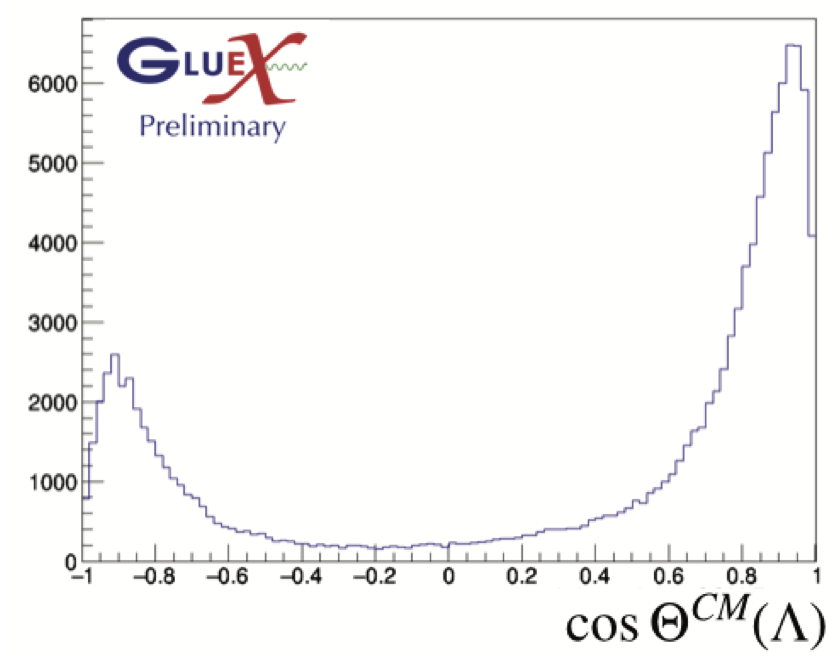}
    \includegraphics[height=3.7cm]{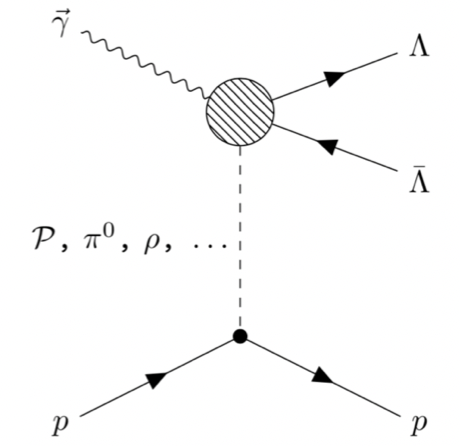}
    \includegraphics[height=3.7cm]{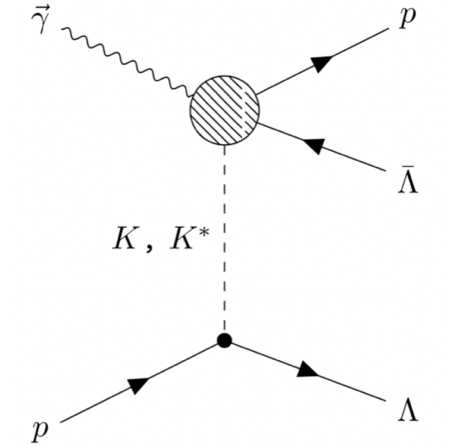}
    \caption{The left panel shows the $\cos\theta$ distribution of $\Lambda$ in the center-of-mass frame for the reaction $\gamma p \rightarrow \Lambda\bar\Lambda p$. The forward and backward peaks can be explained by the two possible production mechanisms shown to the right of it. The middle panel shows the $\Lambda\bar\Lambda$ pair being produced at the top vertex recoiling against a proton. The right panel depicts a process in which strangeness is exchanged and a $p\bar\Lambda$ pair is produced at the top vertex recoiling against a $\Lambda$. All taken from Ref.~\cite{Li:2019rts}.}
    \label{fig:LLprod}
\end{figure}

We focus here on $\Lambda\bar\Lambda$, which we measure in the reaction $\gamma p \rightarrow \Lambda\bar\Lambda p\rightarrow \{\pi^-p\}\{\pi^+\bar p\}p$. Little theoretical work has been done so far to describe the reaction, so a first focus is on the production mechanisms. Studying angular distributions of $\Lambda$ (Fig.~\ref{fig:LLprod} left), $\bar\Lambda$ and $p$ in the centre-of-mass frame reveals that the reaction process seems to be dominated by t-channel exchanges in which the $\Lambda\bar\Lambda$ pair is produced recoiling against the proton (Fig.~\ref{fig:LLprod} middle). However, there is also a sizeable contribution of events where $p\bar\Lambda$ are produced as pair recoiling via strangeness exchange against a $\Lambda$ (Fig.~\ref{fig:LLprod} right). There is no indication of production of a $p\Lambda$ pair recoiling against $\bar\Lambda$.
These two reaction mechanisms are intriguing as they could proceed through intermediate mesons. In order to study this further, the differential cross-sections $d\sigma/d\textbf{IM}(\Lambda\bar\Lambda)$ and $d\sigma/d\textbf{IM}(p\bar\Lambda)$ are measured for the respective systems. Preliminary results are shown in Fig.~\ref{fig:dsig_dMLL}.
\begin{figure}
    \centering
    \begin{minipage}[t]{0.35\textwidth}
    \centering
    \includegraphics[height=4cm]{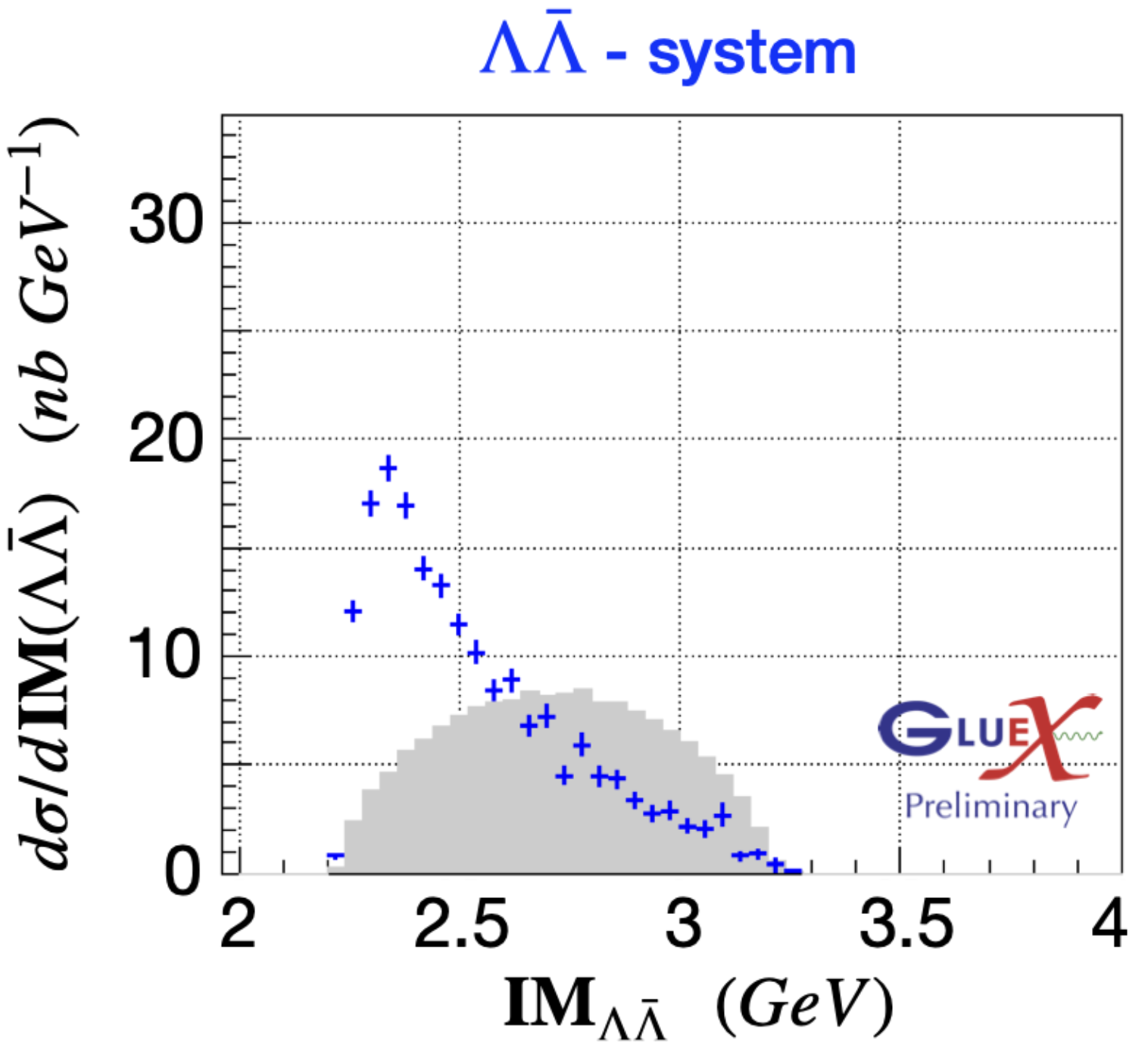}
    \label{fig:dsig_dMLL}
    \end{minipage}
    \begin{minipage}[t]{0.64\textwidth}
    \centering
    \sidecaption
    \includegraphics[height=4cm]{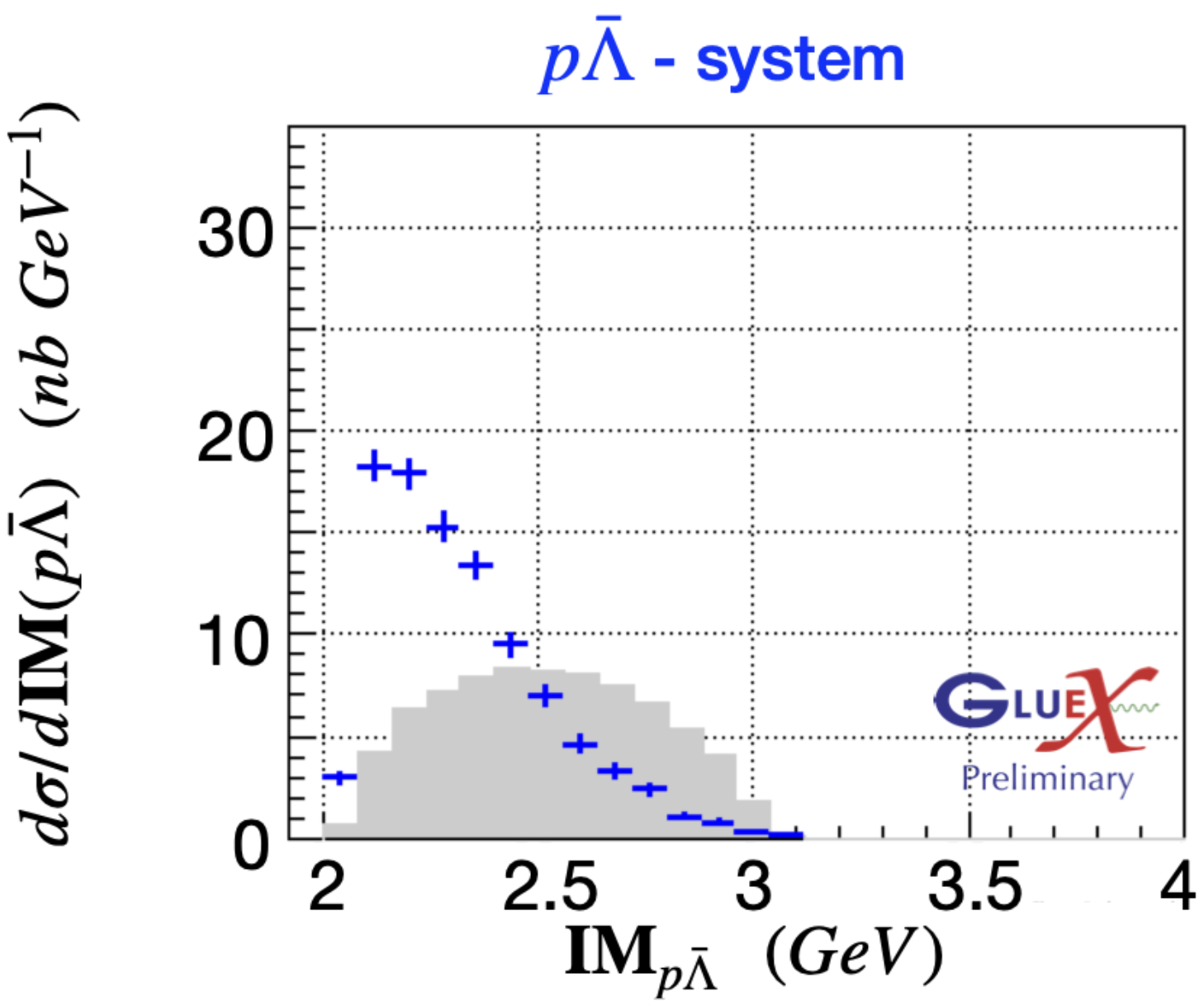}
    \caption{Preliminary differential cross-sections for the $\Lambda\bar\Lambda$ (left) and the $p\bar\Lambda$ (right) systems. The grey areas indicate the distribution expected for production uniform in phase-space.}
    \label{fig:dsig_dMLL}
    \end{minipage}
\end{figure}
The data is shown in blue and the grey area indicates the expected distribution for events produced uniformly in phase-space. Interesting enhancements at threshold are visible in both cases and studies are ongoing to investigate their origin.\par
Further measurements are planned for this reaction. The $\Lambda$, being self-analyzing due to its weak decay, offers the opportunity to measure the polarization of the produced baryon pair. In addition, the linearly polarized photon beam of GlueX can be used to study the beam asymmetry $\Sigma$, which contains information about the nature of the exchanged particle in the t-channel exchange (c.f. Sec.~\ref{sec:Sigma}).

\section{$\Lambda(1405)$ line shape measurements}\label{sec:L1405}
The $\Lambda(1405)$ is arguably one of the most interesting hyperons. Although its existence is long established beyond any doubt, it is still controversial in its nature. Many current analyses support a two-pole structure. For a recent review of the $\Lambda(1405)$ see e.g.~Ref.~\cite{Mai:2020ltx}. The Particle Data Group include a review of the pole structure of the $\Lambda(1405)$ (Chapter 83 in Ref.~\cite{ParticleDataGroup:2020ssz}) and recently added the $\Lambda(1380)$, the lower of the two poles, to its particle listings. One key measurement to help with the interpretation of the $\Lambda(1405)$ is the precise measurement of its line shape. Due to its good energy and momentum resolutions, GlueX is well suited for this task.\par
\begin{figure}[h]
    \centering
    \includegraphics[width=0.49\linewidth]{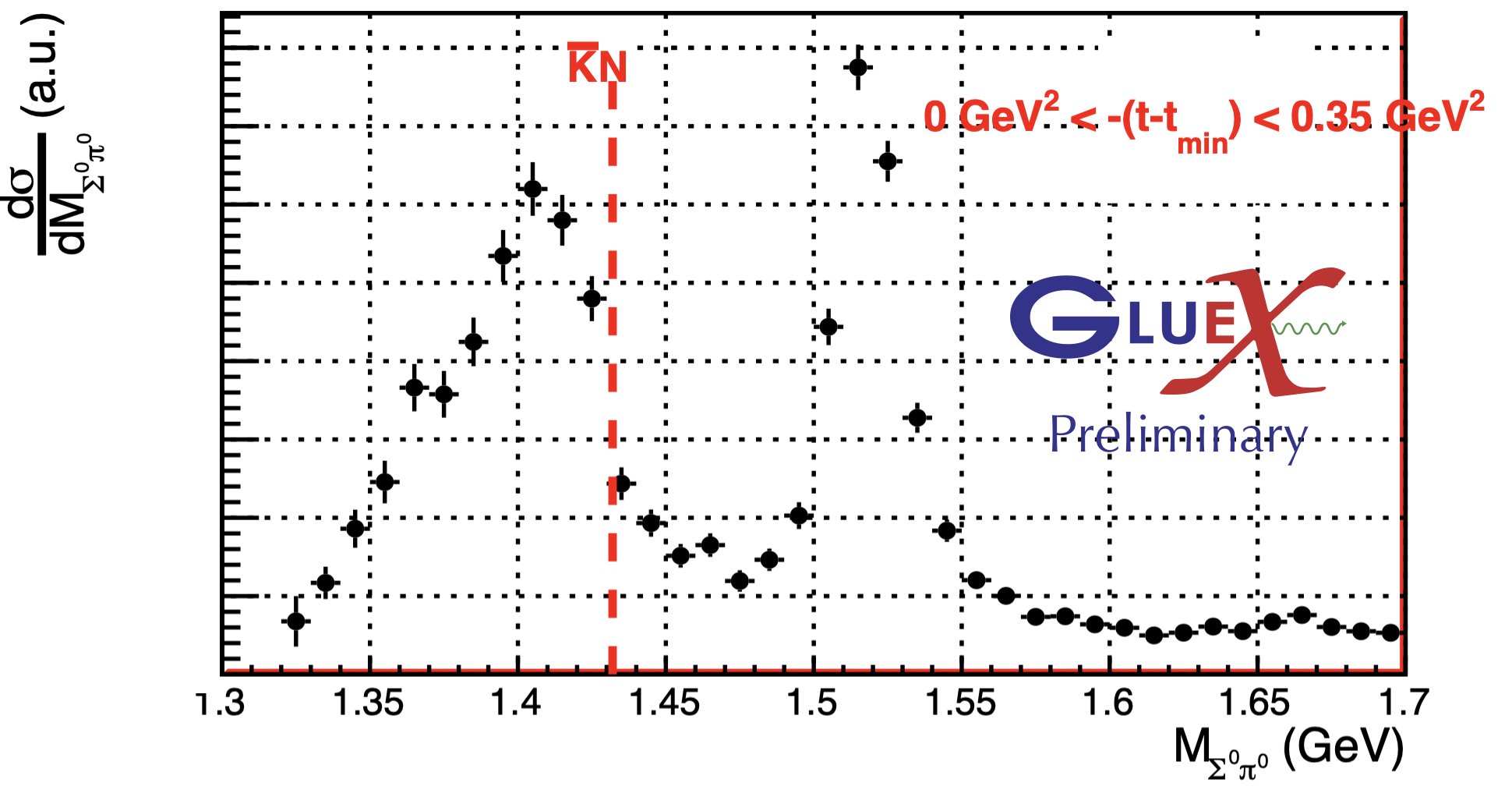}
    \includegraphics[width=0.49\linewidth]{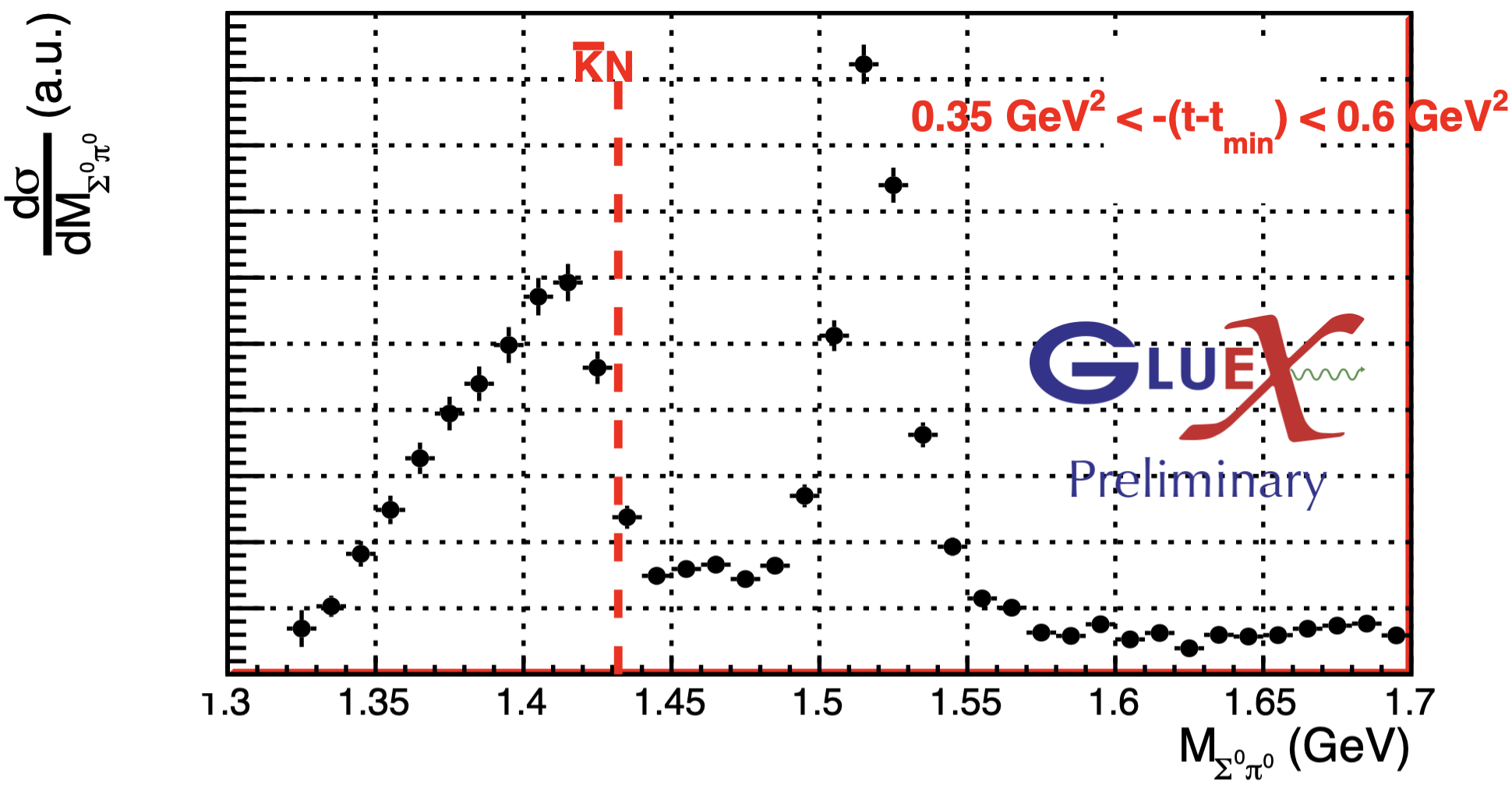}
    \sidecaption
    \includegraphics[width=0.49\linewidth]{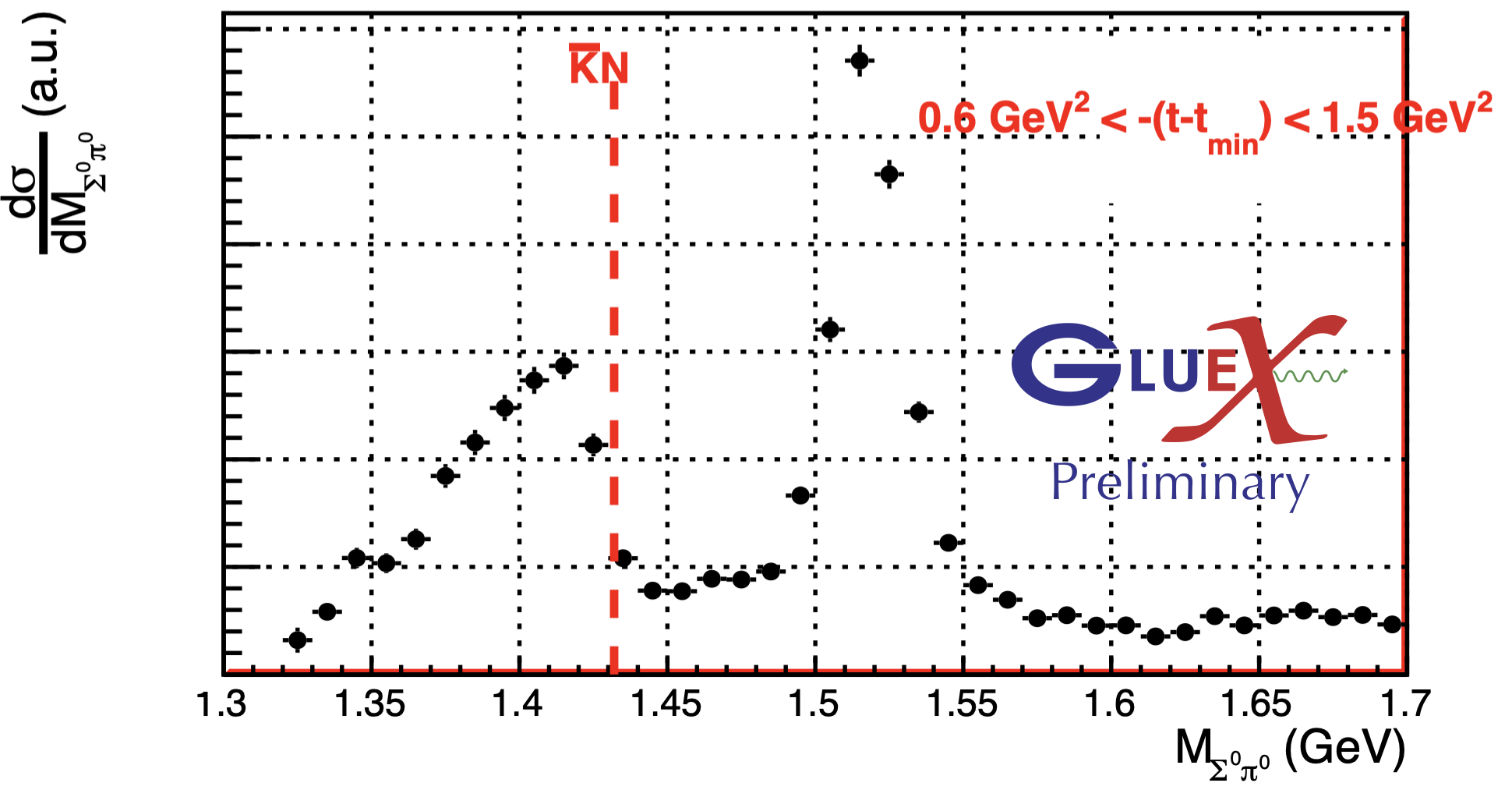}
    \caption{Preliminary results for the differential cross-section measurement of $\gamma p\rightarrow K^+\Sigma^0\pi^0$ in three bins of momentum transfer $-t=-(p_\text{beam}-p_{K^+})^2$. The $\Lambda(1405)$ and $\Lambda(1520)$ are clearly identifiable as two peaking structures. The $\Lambda(1405)$ shows a very much, non-Breit-Wigner shape with a sharp drop at the $\bar KN$ threshold.}
    \label{fig:L1405}
\end{figure}

The reaction of interest is $\gamma p\rightarrow K^+\Lambda(1405)\rightarrow K^+\Sigma^0\pi^0$ with $\Sigma^0\rightarrow\Lambda\gamma\rightarrow p\pi^-\gamma$ and $\pi^0\rightarrow\gamma\gamma$. The $\Sigma^0\pi^0$ decay of the $\Lambda(1405)$ is ideally suited to study the resonance as it is free from $\Sigma(1385)$ background since it is isospin $I=0$. The measurement is done in the photon energy range $E_\gamma=\SIrange{6.5}{11.6}{\GeV}$. In total, about 13,350 events where identified in the mass region of the $\Lambda(1405)$. The mass resolution for these events is less than \SI{5}{\MeV}. The data was split in three bins of momentum transfer $-t=-(p_\text{beam}-p_{K^+})^2$. The preliminary results are shown in Fig.~\ref{fig:L1405}.
One can easily identify the $\Lambda(1405)$ and $\Lambda(1520)$ as two peaking structures. The $\Lambda(1405)$ clearly shows a non-Breit-Wigner shape with a sharp drop at the $\bar KN$ threshold. It seems that the line shape of the $\Lambda(1405)$ shows a slight t-dependence, which is consistent with a two-pole structure. Investigations into the extraction of the number and positions of poles are currently ongoing.\par
A comprehensive report on preliminary results by GlueX can be found elsewhere in these proceedings~\cite{Nilanga:1405}. 

\section{$\Lambda(1520)$ spin-density matrix elemements and cross-sections}\label{sec:L1520}
Another excited $\Lambda$ hyperon under investigation at GlueX is the $\Lambda(1520)$. This $J^P=\frac{3}{2}^-$ state is observed in its decay to $K^-p$. Therefore, the reaction of interest is $\gamma p \rightarrow K^+\Lambda(1520)\rightarrow K^+K^-p$. The $\Lambda(1520)$ is a very well-established state and a better understanding of its production process will be very helpful for other analyses at GlueX where strangeness exchange in the t-channel is important.\par
\begin{figure}[h]
    \centering
    \includegraphics[width=\linewidth]{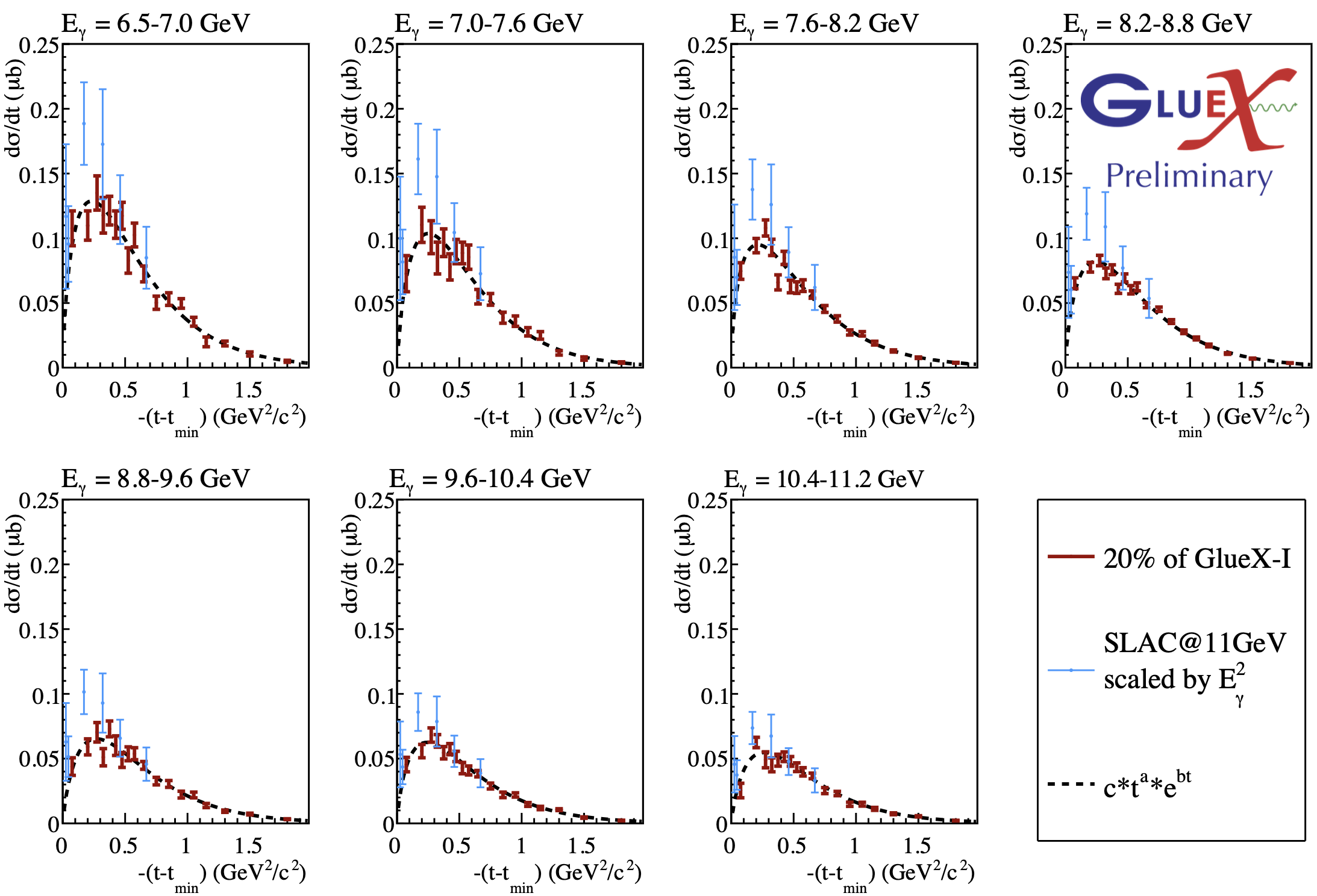}
    \caption{Preliminary differential cross-section measurements for $\Lambda(1520)$ phoptoproduction in seven bins of beam energy and 18 bins of momentum transfer $-t=-(p_\text{beam}-p_{K^+})^2$ are shown in red. The blue data points show a previous measurements performed at SLAC~\cite{Boyarski:1970yc} at $E_\gamma=\SI{11}{\GeV}$ and scaled down by $E_\gamma^2$. The black dashed line is the fit used to integrate the differential cross-section data to get the total cross-section.}
    \label{fig:L1520}
\end{figure}

GlueX has published a paper on spin-density matrix elements (SDMEs) in this reaction~\cite{GlueX:2021pcl}. SDMEs parametrize the angular decay distribution of the $K^-$ in the $\Lambda(1520)$ rest frame. Studying them allows to draw conclusions about the production mechanisms involved in the reaction. More specifically, assuming t-channel exchange as the dominant production mode, one can study the naturality $\eta=P(-1)^J$ of the exchanged particle with quantum numbers $J^P$. Utilizing the linear polarization of our photon beam, we measured six polarized SDMEs for the first time, in addition we also extracted three independent unpolarized SDMEs. We found, that the production of $\Lambda(1520)$ between $E_\gamma=\SIrange{8.2}{8.8}{\GeV}$ is dominated by natural exchanges, i.e. vector or tensor meson exchange. In order to make more precise statements on the exact nature of the exchanged mesons, additional information is needed. For this reason, we are currently working on the extraction of the differential cross-section $\frac{d\sigma}{dt}$. We study $20\%$ of the GlueX-I data and measure $\frac{d\sigma}{dt}$ in 18 bins in momentum transfer $-t=-(p_\text{beam}-p_{K^+})^2$ and seven bins in photon beam energy $E_\gamma$. Our preliminary results are shown in Fig.~\ref{fig:L1520} in red and compared to the blue data points measured by an experiment at SLAC~\cite{Boyarski:1970yc}.
SLAC performed their measurements at $E_\gamma=\SI{11}{\GeV}$ but in their paper they claim an expected dependency of $1/E_\gamma^2$ for the total cross-section. For this reason, and to compare to our results, we scaled the SLAC data points according to this factor to our energy bins and plotted them in each $E_\gamma$ bin. We see good agreement between the SLAC data and our data. The agreement is better for higher photon beam energies and gets slightly worse towards lower $E_\gamma$, indicating that the scaling factor quoted by SLAC might not be accurate at low energies.\par
\begin{figure}
    \centering
    \sidecaption
    \includegraphics[width=0.7\linewidth]{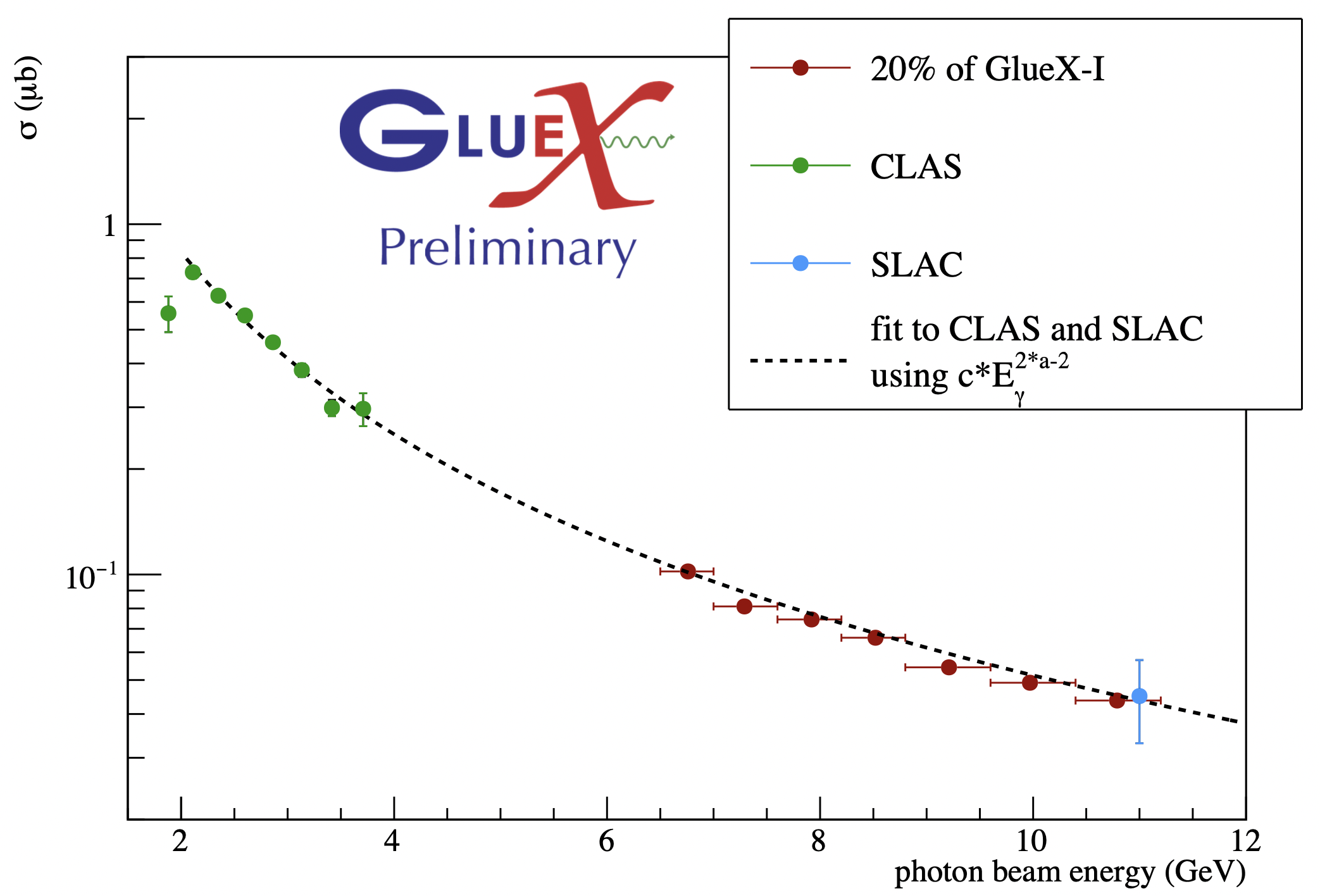}
    \caption{Preliminary total cross-section results are show in red. The green data at low energies comes from the CLAS experiment~\cite{CLAS:2013rxx} the data point at \SI{11}{\GeV} is from SLAC~\cite{Boyarski:1970yc}. The black dashed curve is a phenomenological fit of the form $c*E_\gamma^{2a-2}$ to the CLAS and SLAC data. Our preliminary data, which was not fitted, agrees well with the black curve.}
    \label{fig:L1520total}
\end{figure}

In order to measure total cross-sections, we fit a function with the form $c*t^a*e^{bt}$ to the differential cross-section results and integrate it to large $-t$. The preliminary total cross-section results are shown in Fig.~\ref{fig:L1520total} in red.
The green data points come from CLAS~\cite{CLAS:2013rxx} and the blue data point at \SI{11}{\GeV} was measured by SLAC~\cite{Boyarski:1970yc}. In order to compare the results to our preliminary data, we fitted the CLAS and SLAC data with $c*E_\gamma^{2a-2}$ (black dashed curve), which we found to better describe the data than $c*E_\gamma^{-2}$, as suggested by SLAC. The preliminary GlueX results, which were not part of the fit, agree very well with the fit. We were also able to take some data at lower photon beam energies during our 2018 data taking campaign. This data which is yet to be analysed will cover most of the gap between the CLAS data and our preliminary results.

\section{$\Sigma^0$ hyperon beam asymmetry}\label{sec:Sigma}
\begin{figure}[h]
    \centering
    \sidecaption
    \includegraphics[width=0.5\linewidth]{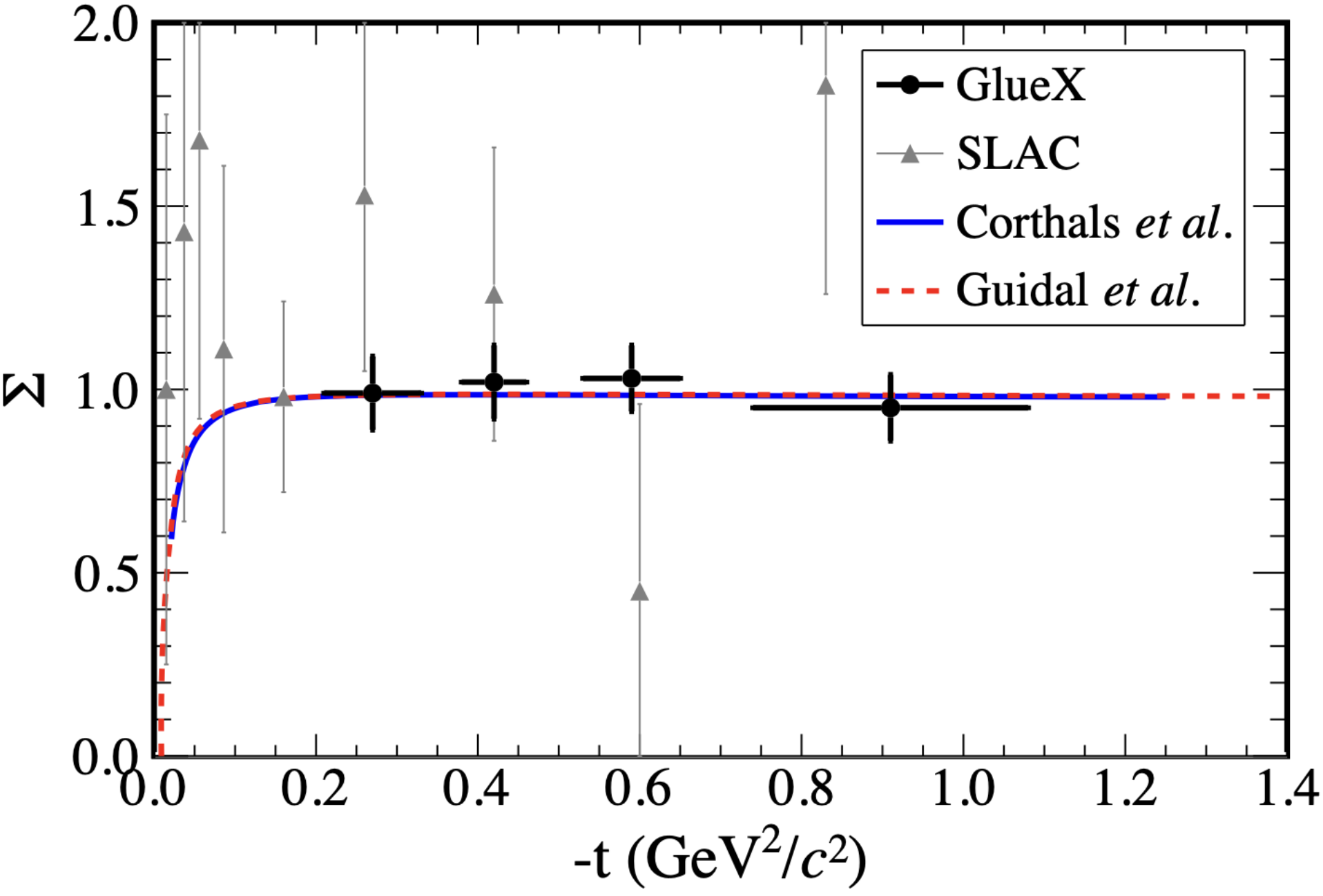}
    \caption{GlueX results for the beam asymmetry $\Sigma$ in bins of momentum transfer $-t=-(p_\text{beam}-p_{K^+})^2$ for $\Sigma^0$ photoproduction based on about $20\%$ of GlueX-I are shown in black. The grey triangles show previous measurements carried out at SLAC~\cite{PhysRevD.20.1553} at $E_\gamma=\SI{16}{\GeV}$. The data is compared to two model calculations by Corthals {\it et al.}~\cite{PhysRevC.73.045207,PhysRevC.75.045204} and Guidal {\it et al.}~\cite{GUIDAL1997645}, both predicting a beam asymmetry of almost one over the measured t-range. Taken from Ref.~\cite{GlueX:2020qat}.}
    \label{fig:Sigma_BA}
\end{figure}

In addition to the various $\Lambda$ hyperons, GlueX also studied the ground state $\Sigma^0$. It is observed in the reaction $\gamma p\rightarrow K^+\Sigma^0$ with the subsequent decays $\Sigma^0\rightarrow\Lambda\gamma\rightarrow p\pi^-\gamma$. In order to learn more about how the $\Sigma^0$ is produced, we have measured the beam asymmetry $\Sigma$, which is accessible thanks to GlueX's linearly polarized photon beam, in bins of momentum transfer $-t=-(p_\text{beam}-p_{K^+})^2$. $\Sigma$ can be measured as a counting rate asymmetry of $\Sigma^0$ events between two orthogonal settings of the polarization axis of the beam. The results are published in Ref.~\cite{GlueX:2020qat} and shown in Fig.~\ref{fig:Sigma_BA}.
The black points show the GlueX results, which are consistent with one over the whole measured range in $-t$. The grey triangles are previous measurements performed by SLAC~\cite{PhysRevD.20.1553} at $E_\gamma=\SI{16}{\GeV}$. They have large uncertainties but generally agree with the GlueX data. Also shown are two model calculations by Corthals {\it et al.}~\cite{PhysRevC.73.045207,PhysRevC.75.045204} and Guidal {\it et al.}~\cite{GUIDAL1997645}. Both predict a beam asymmetry close to one over most of the range in momentum transfer. This large beam asymmetry implies natural parity exchange, similar to what is also observed in the $\Lambda(1520)$ measurement.

\section{Cascade production}\label{sec:Cascade}
GlueX can not only provide important data for single strangeness hyperons but also observes various doubly strange hyperons. This is especially interesting since there seems to be a large discrepancy between the number of expected and number of known states with strangeness $S=-2$. GlueX can provide important data to help identify new states in this sector. \par
Cascade baryons are expected to be produced slightly differently compared to $\Lambda$ or $\Sigma$ hyperons. Instead of being produced in a direct t-channel exchange, which can only generate one unit of strangeness, cascade baryon production has to proceed through a decay of a high mass hyperon with $S=-1$, which decays into a kaon and the cascade hyperon adding another unit of strangeness. Therefore, we expect to observe the reactions through $\gamma p\rightarrow K^+(\Lambda^*/\Sigma^*)\rightarrow K^+K^{+/0}\Xi^{-/0}$.\par
\begin{figure}
    \centering
    \begin{minipage}[t]{0.54\textwidth}
        \centering
        \includegraphics[width=\linewidth]{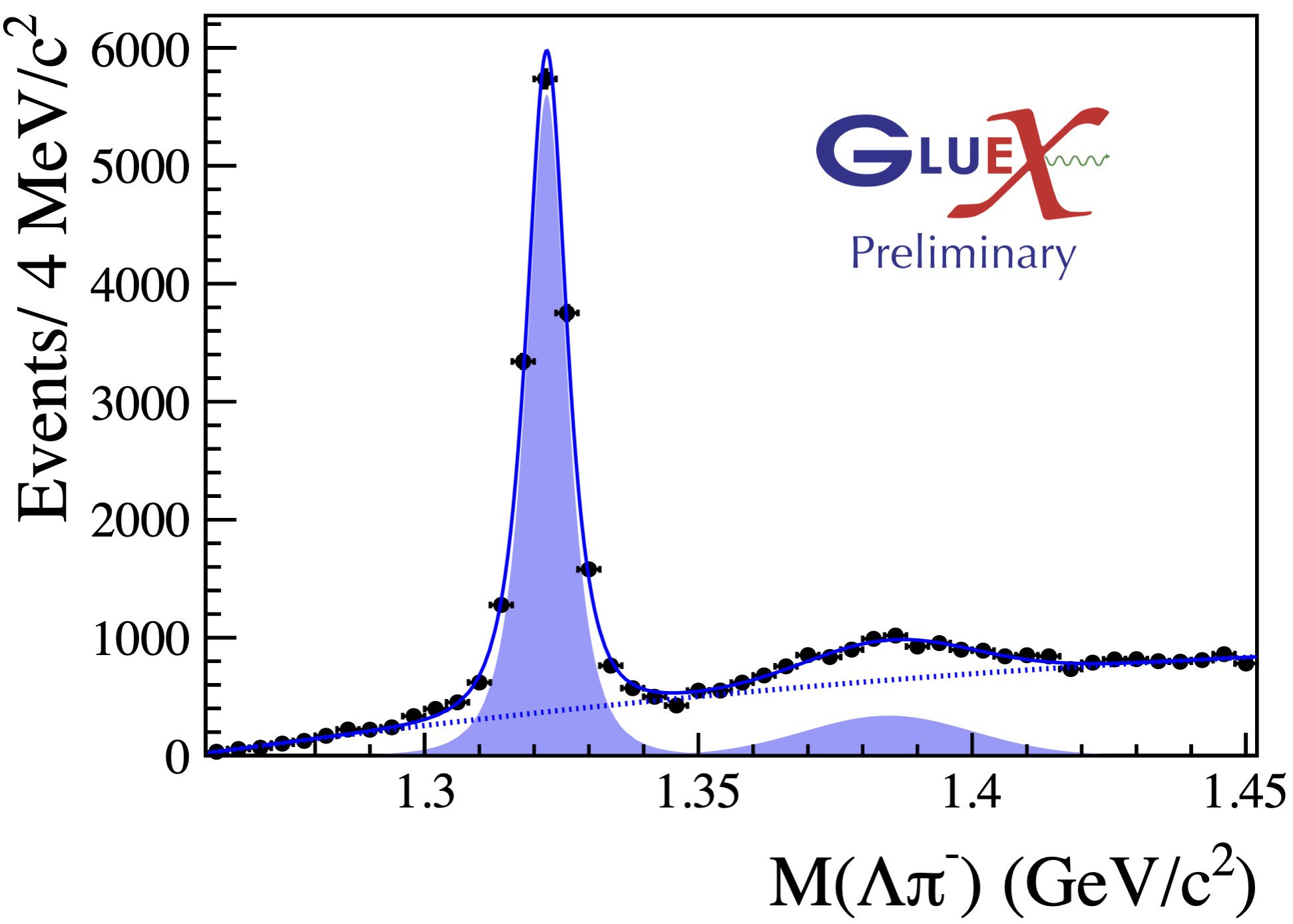}
        \caption{Preliminary invariant mass spectrum of the $\Lambda\pi^-$ subsystem in the reaction $\gamma p\rightarrow K^+K^+\Xi^-\rightarrow K^+K^+\pi^-\Lambda\rightarrow K^+K^+\pi^-\pi^-p$. The $\Xi^-(1320)$ can be easily identified as a narrow peak at the expected mass. The bump at higher masses is due to the $\Sigma(1385)$ with a miss-identified pion.}
        \label{fig:Xi1320}
    \end{minipage}
    \hspace{0.3cm}
    \begin{minipage}[t]{0.42\textwidth}
        \centering
        \includegraphics[width=\linewidth]{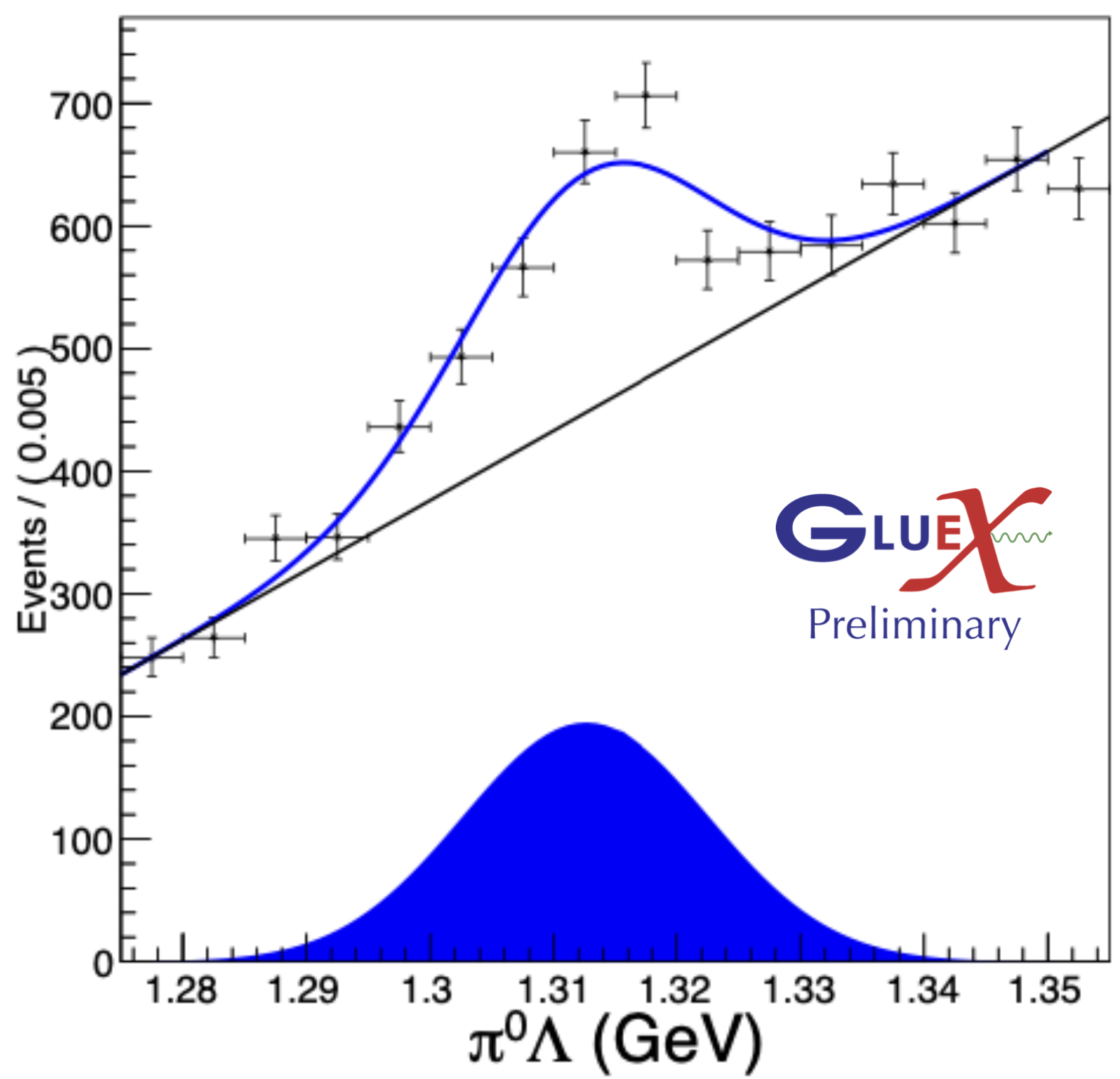}
        \caption{Preliminary invariant mass spectrum of the $\pi^0\Lambda$ subsystem in the reaction $\gamma p\rightarrow K^+K_s^0\Xi^0\rightarrow K^+\{\pi^+\pi^-\}\pi^0\Lambda\rightarrow K^+\{\pi^+\pi^-\}\gamma\gamma\pi^-p$. The $\Xi^0(1315)$ can be identified as a peak at the expected mass on top of some moderate background.}
        \label{fig:Xi1315}
    \end{minipage}
\end{figure}

The ground state cascades decay predominantly into $\pi\Lambda$. Therefore, in the search for the charged ground state cascade the reaction of interest is $\gamma p\rightarrow K^+K^+\Xi^-\rightarrow K^+K^+\pi^-\Lambda\rightarrow K^+K^+\pi^-\pi^-p$. Figure~\ref{fig:Xi1320} shows the preliminary invariant mass spectrum of the $\Lambda\pi^-$ system. One can clearly identify a strong and narrow signal peak at the mass of the $\Xi^-(1320)$.
In order to observe the neutral ground state cascade $\Xi^0(1315)$ we study the reaction $\gamma p\rightarrow K^+K_s^0\Xi^0\rightarrow K^+\{\pi^+\pi^-\}\pi^0\Lambda\rightarrow K^+\{\pi^+\pi^-\}\gamma\gamma\pi^-p$. This final state has a higher multiplicity and contains photons in the final state. Nevertheless, GlueX is able to identify the $\Xi(1315)$ in the $\pi^0\Lambda$ invariant mass spectrum as a peak on top of some background, as shown in Fig.~\ref{fig:Xi1315}.\par
\begin{figure}
    \begin{minipage}[t]{0.49\textwidth}
        \centering
        \includegraphics[width=\linewidth]{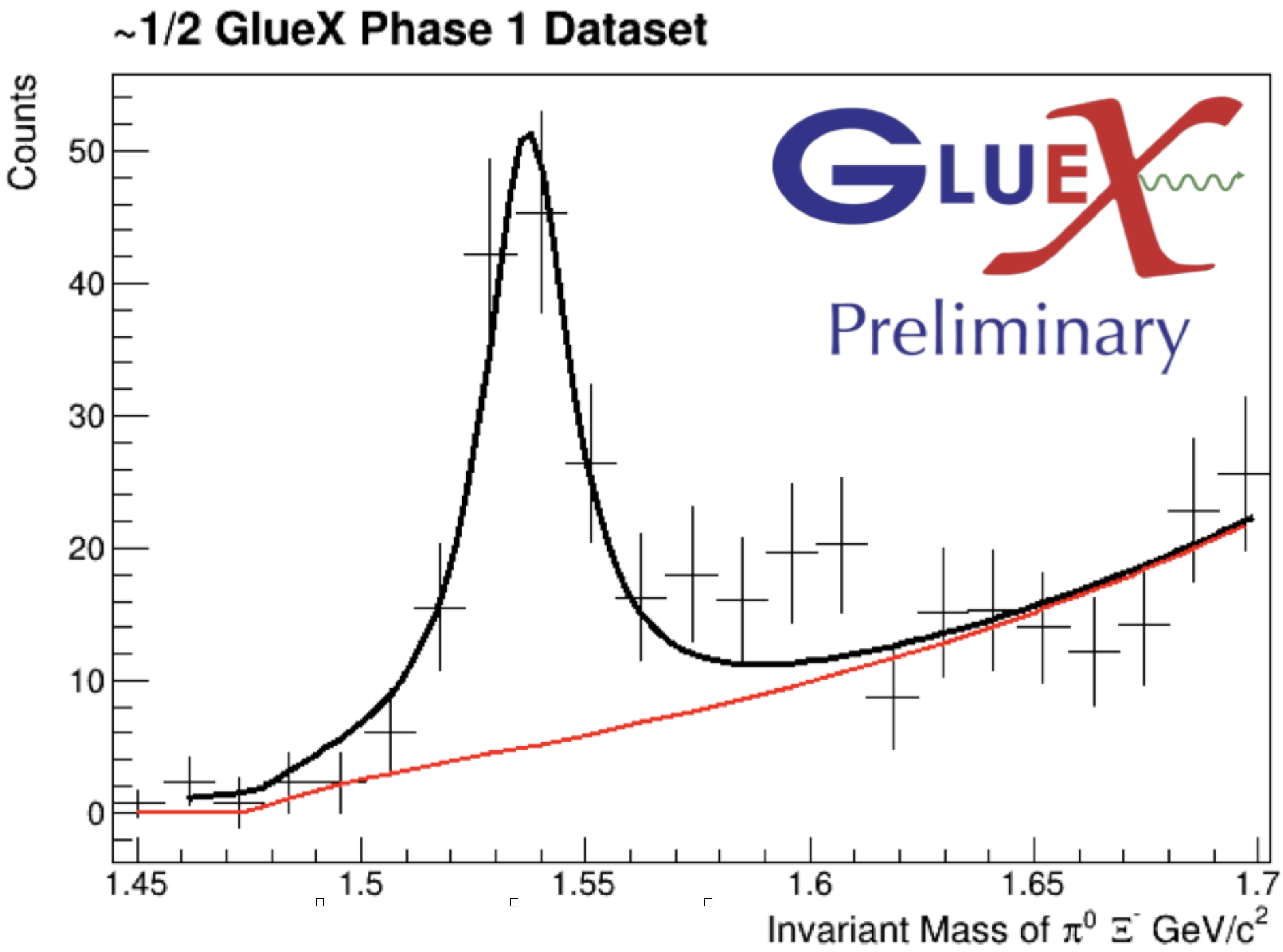}
        \caption{Preliminary invariant mass spectrum of the $\pi^0\Xi^-$ subsystem in the reaction $\gamma p\rightarrow K^+K^+\Xi^{*-}\rightarrow K^+K^+\pi^0\Xi^-\rightarrow K^+K^+\gamma\gamma\pi^-\Lambda\rightarrow K^+K^+\gamma\gamma\pi^-\pi^-p$. The $\Xi^{*-}(1530)$ can be clearly identified as a peak at the expected mass.}
        \label{fig:Xi1530}
    \end{minipage}
    \hspace{0.3cm}
    \begin{minipage}[t]{0.47\textwidth}
        \centering
        \includegraphics[width=\linewidth]{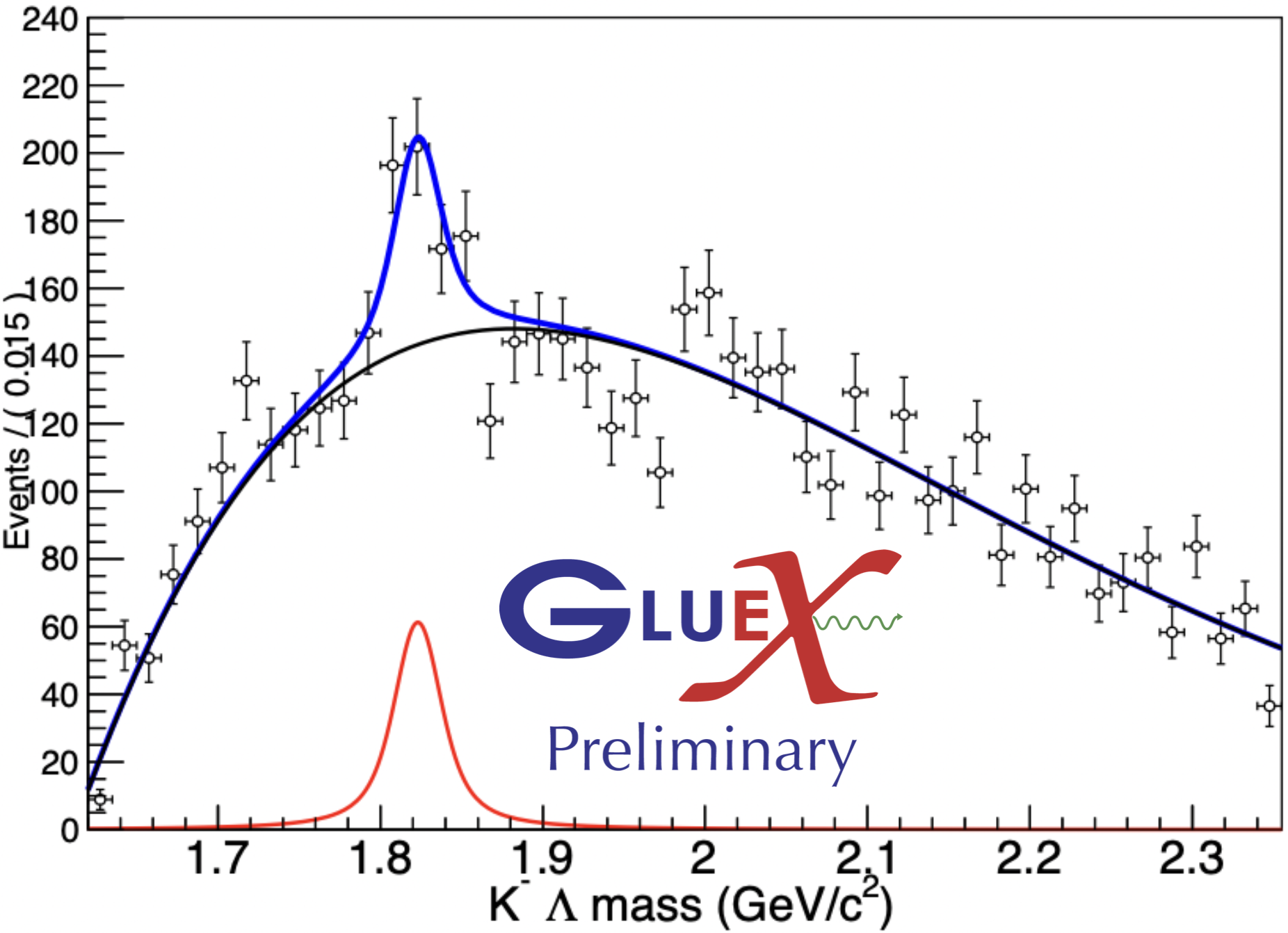}
        \caption{Preliminary invariant mass spectrum of the $K^-\Lambda$ subsystem in the reaction $\gamma p\rightarrow K^+K^+\Xi^{*-}\rightarrow K^+K^+K^-\Lambda\rightarrow K^+K^+K^-\pi^-p$ for $E_\gamma=\SIrange{7.4}{8.0}{\GeV}$. The $\Xi^{*-}(1820)$ can be identified as a peak on some background distribution.}
        \label{fig:Xi1820}
    \end{minipage}
\end{figure}

The excited $\Xi^{*-}(1530)$ is known to decay into $\pi^0\Xi^-$. In order to search for it, the reaction $\gamma p\rightarrow K^+K^+\Xi^{*-}\rightarrow K^+K^+\pi^0\Xi^-\rightarrow K^+K^+\gamma\gamma\pi^-\Lambda\rightarrow K^+K^+\gamma\gamma\pi^-\pi^-p$ is studied. This is again a final state with both charged and neutral particles and high multiplicity. Nevertheless, studying about $50\%$ of the GlueX-I data set reveals a peaking structure at the expected mass, as shown in Fig.~\ref{fig:Xi1530}.\par
The highest mass excited cascade we have identified so far is the $\Xi^{*-}(1820)$, which was seen for the first time in photoproduction by GlueX. We observe it in the reaction $\gamma p\rightarrow K^+K^+\Xi^{*-}\rightarrow K^+K^+K^-\Lambda\rightarrow K^+K^+K^-\pi^-p$. It can be identified as a peak in the preliminary $K^-\Lambda$ invariant mass spectrum as shown in Fig.~\ref{fig:Xi1820}. It is noteworthy that it is the only excited cascade hyperon that clearly shows up as peak in the broad range of invariant mass shown in the plot.\par
Despite experimental challenges like mixed final states and high multiplicities GlueX can clearly identify a range of excited and ground state cascade hyperons. Going forward we will work towards measurements of physics observables for these reactions.

\section{Summary and outlook}\label{sec:summary}
The GlueX experiment at Jefferson Lab is an exceptional photoproduction facility which, thanks to its comprehensive detector system, is able to detect a wide range of charged and neutral final state particles. This allows GlueX to measure many strangeness related reactions. GlueX establishes the production mechanisms for a wide range of different hyperons and, thereby, paves the way for future partial wave analyses which rely on good knowledge of the involved reactions. In addition, GlueX can search for and identify many (excited) hyperons with strangeness $S=-2$. This is very encouraging for future searches for new strange particles. \par

In addition to the hyperon measurements reported here, GlueX also has a range of active strangeness analyses in the meson sector. We observe many excited kaons and other mesons decaying to final states containing strangeness. With the addition of the DIRC detector to our GlueX spectrometer setup, we expect improved kaon identification in our GlueX-II data. This will enable us to continue to make important contributions to strange hadron spectroscopy. The planned measurements comprise, but are not limited to, searches for strangeonia, the $Y(2175)$ and strange hybrid mesons. In addition, we will continue our program of searches for hyperons. \par

We would like to acknowledge the outstanding efforts of the staff of the Accelerator and the Physics Divisions at Jefferson Lab that made the experiment possible. This work was supported in part by the UK Science and Technology Facilities Council. This material is based upon work supported by the U.S. Department of Energy, Office of Science, Office of Nuclear Physics under contract DE-AC05-06OR23177.
For a complete list of founding bodies and acknowledgements please consult \url{gluex.org/thanks}.

\bibliography{refs}
\end{document}